\documentclass{article}
\usepackage{amssymb}
\usepackage{amsmath}

\newtheorem{theorem}{Theorem}

\newtheorem{definition}[theorem]{Definition}

\begin{document}

\title{Coisotropic deformations of algebraic  varieties and
integrable systems}
\author{B.G.Konopelchenko* and G.Ortenzi** \\
%EndAName
\\
*Dipartimento di Fisica, Universit\`{a} del Salento \\
and INFN, Sezione di Lecce, 73100 Lecce, Italy \\
**Dipartimento di Matematica Pura ed Applicazioni, \\
Universit\`{a} di Milano Bicocca, 20125 Milano, Italy}
\maketitle

\begin{abstract}
Coisotropic deformations of algebraic varieties  
are defined as those for which an ideal of the deformed variety is a Poisson ideal. It is shown that coisotropic deformations of sets of intersection points
of plane quadrics, cubics and space algebraic curves are governed, in particular, by the dKP, WDVV,
 dVN, d2DTL equations and other integrable hydrodynamical type systems. Particular attention is paid to the study of 
two- and three-dimensional deformations of elliptic curves. Problem of an appropriate choice of 
Poisson structure is discussed.
\end{abstract}

\section{Introduction}

\ \ Algebraic varieties  (curves etc.) and their deformations are important
ingredients in various branches of mathematics and mathematical physics.
Theory of integrable nonlinear differential equations was, probably, the
most active area in recent years where these objects and their properties
have been intensively studied. \ Two best known examples of such study are given
by the theory of the finite-gap solutions and the theory of the \ Whitham
equations \cite{NMPZ}-\cite{K2}. The problem of characterization and classification of
integrable deformations of algebraic curves \ has attracted a particular
interest. \ In the papers \cite{K1,K2} Krichever formulated a general theory of
hierarchies of integrable equations of hydrodynamical type on a Riemann
surfaces of arbitrary genus arising in the Whitham averaging method.

An alternative approach for determining and classifying \ the so-called
quasiclassical deformations of algebraic curves has been proposed in \cite{KK,KMA,KKMAM}.
This \ approach revealed a deep connection between the structure of possible
deformations of algebraic curves and their basic algebraic properties like
the Galois group \cite{KMA,KKMAM}. Quite different method of study of the Whitham
equations has been discussed recently by Magri \cite{M}. Deformations studied in
the papers \cite{KK,KMA,KKMAM,M} are all two-dimensional, i.e. parametrized by two
variables.

In the present paper we will introduce and study a novel class of
deformations of algebraic curves, surfaces and algebraic varieties, the
class of coisotropic deformations. \ A concept of coisotropic deformations
of associative algebras has been formulated recently in the papers \cite{KM1,KM2,KM3}.
Notions of coisotropic submanifold and Poisson ideal are the basic one for this approach. Here
we will show that essentially the same idea provides us with a simple and
transparent way to define and describe coisotropic deformations of algebraic
varieties in affine spaces. Namely, coisotropic deformations of algebraic variety are those
for which an ideal of deformed variety is a Poisson ideal, i.e. it is closed with respect to 
Poisson bracket. We will consider simple examples of algebraic varieties such as sets of
 intersection points of
algebraic curves, algebraic curves and hypersurfaces. It is shown that the coisotropic
 deformations of these objects are
governed by systems of differential equations of hydrodynamical type which in
particular cases coincide with well-known integrable systems like the dispersionless 
KP equation, WDVV equation and
dispersionless 2DTL equation.

\ We will concentrate on the study of the coisotropic deformations on plane
and three-dimensional  quadrics and cubics. Particular attention will be
paid to the study of the three-dimensional coisotropic deformations of
elliptic curves. The problem of choice of the Poisson structure is
discussed too. It is shown that such a choice is crucial for construction of
nontrivial coisotropic deformations.

The paper is organized as follows. General formulation of coisotropic
deformations of algebraic varieties is discussed in section 2. Coisotropic
deformations of the sets of intersection points of algebraic curves on the
plane are considered in section 3. It is shown that two particular classes
of deformations are governed by the stationary dKP equation and WDVV
equation. Section 4 is devoted to deformations of plane cubics. 
Deformations of the space curves are discussed in the next section 5. The dKP
equation and the dispersionless Veselov-Novikov equation govern coisotropic
deformations of the special space curves. 
Two-dimensional coisotropic deformations of the elliptic curve are studied in section 6.  The problem
of choice of the Poisson structure is discussed here too. 
In section 7 we consider three-dimensional deformations of elliptic curves.
Deformations of curves and hyperplanes in $R^{4}$ described by the Boyer-Finley-d2DTL
equation and heavenly equation are presented in section 8.

\section{Coisotropic deformations of algebraic varieties}

The notion of coisotropic submanifold is a basic ingredient in the formulation
and description of coisotropic deformations of associative \ algebras \
studied in \cite{KM1,KM2}. Coisotropic submanifold $\Gamma $ is a submanifold in 
$R^{2n}$ endowed with the Poisson bracket \{ , \} such that $\Gamma
_{\intercal }\subset \Gamma $ where $\Gamma _{\intercal }$ denotes a
skew-orthogonal complement of $\Gamma $ in $R^{2n}$ (see e.g. \cite{Wein,B}).
Coisotropic submanifold can be defined as the zero locus $\Gamma $ for the
set of functions $f_{j}(y),$ i.e.

\begin{equation}
\label{def1}
f_{j}(y)=0,\quad j=1,...,m
\end{equation}
such that

\begin{equation}
\label{def2}
\left\{ f_{j}(y),f_{k}(y)\right\} |_{ \Gamma }=0,\quad j,k=1,...,m
\end{equation}
where $y_{1},...,y_{n}$ are local coordinates in $R^{2n}$. \ The definition
(\ref{def1}), (\ref{def2}) is equivalent to the condition that all the Hamiltonian fields
generated by $f_{j}(y)$ ate tangent to $\Gamma $ or to the condition that
the ideal $J=\langle f_{j}\rangle $ generated by the functions $f_{j}(y)$ is
closed \ $\left\{ J,J\right\} \subset J$, i.e. it is a Poisson ideal \cite{Wein,B}. For associative algebras
the functions (\ref{def1}) are of the form \cite{KM1}

\begin{equation} 
\label{assalg}
f_{jk}=-p_{j}p_{k}+\sum_{l=1}^{n}C_{jk}^{l}(x)p_{l},\quad j,k=1,..,n
\end{equation}
where $p_{j},x_{j}(j=1,..,n)$ are canonical Darboux coordinates in $R^{2n}$.
The conditions (\ref{def2}) define the coisotropic deformations of the structure
constants $C_{jk}^{l}(x)$ of a associative algebra in a given basis. It was
shown in \cite{KM2} that this approach is applicable to the other algebraic
structures like the Jordan triples. In geometrical terms equations (\ref{def1}), (\ref{assalg})
with fixed $x$ represent a set of special quadrics in $R^{n}$.

Here we will use the notions of Poisson ideal and coisotropic submanifold in a different
setting. It will serve us to define and describe a class of deformations of
algebraic varieties. Thus, let us consider an algebraic variety M in $R^{n}$ defined by
the equations

\begin{equation}
\label{varM}
f_{j}(p_{1},...,p_{n})=0,\quad j=1,..,m
\end{equation}
where $p_{1},...,p_{n}$ are local affine coordinates in $R^{n}$.

To define deformation of this variety 

1) we assume that the coefficients of the polynomials $f_{j}(p_{1},...,p_{n})$
depend on the deformation parameters $x_{1},...,x_{n}$,

2) we embed the variety M into the space $R^{2n}$ equipped with the
Poisson bracket $\left\{ ,\right\} $ and local coordinates $%
p_{1},...,p_{n},x_{1},...,x_{n},$

3) then we consider an ideal $J=\langle f_{j}(p;x)\rangle$ generated by the functions 
$ f_{j}(p;x)$  and require that this ideal is closed
\begin{equation}
\left\{ J , J \right\} \subset J
\end{equation}
or equivalently
\begin{equation}
\label{coisoeqn}
\left\{ f_{j}(p;x),f_{k}(p;x)\right\}\mid_{ \Gamma }=0,\quad j,k=1,..,m
\end{equation}
where $\Gamma $ is the locus of common zeros for the functions $f_{j}(p,x)$,
i.e.
\begin{equation}
\label{subGamma}
\Gamma =\left\{ (p,x) \mid f_{j}(p;x)=0,\ j=1,..,m\right\} .
\end{equation}
In other words we require that the ideal J of the deformed variety M is a Poisson ideal.
\begin{definition}
Deformations of the algebraic variety M defined by equations (\ref{varM}) are called coisotropic if the ideal 
$J=\langle f_{j}(p;x)\rangle$ of the deformed varieties is a Poisson ideal.
\end{definition}

 So, coisotropic deformations of an algebraic variety are those for which coefficients of the functions $f_{j}(p,x)$ are such that the conditions (\ref{coisoeqn}) are satisfied, i.e. the submanifold $\Gamma$ defined by (\ref{subGamma}) is a coisotropic submanifold.

The coisotropy conditions (\ref{coisoeqn}) impose constraints on the coefficients of the
polynomials $f_{j}(p;x)$. We will refer to the corresponding system of
equations for these coefficients as the central system (CS). 
 \ The concrete form of CS depends on the choice of $f_{j}(p;x)$ as well as the form of the
Poisson bracket $\left\{ ,\right\} $. The choice of the Poisson bracket is a
crucial one. For inadequate choice of $\left\{ ,\right\} $ one may have no
nontrivial deformation. The consistency of the Poisson structure $\left\{
,\right\} $ with the polynomials $f_{j}(p;x)$ is an important point of the
approach under consideration.

At $m=n$ the variety M is a set of intersection points of n algebraic curves $
f_{j}(p)=0$ in $R^{n}$ and coisotropic deformations of each of these points
span Lagrangian submanifolds in $R^{2n}$. \ At $m=n-1$ the variety M is a
curve in $R^{n}$ and the conditions (\ref{coisoeqn}) define coisotropic deformations of
this curve and so on. Clearly, the codimension of the algebraic variety
should be greater or equal to two, i.e. $m\geq 2$ in order to be able to
define its coisotropic deformation.

\section{ Coisotropic deformations on the plane. Deformations of the sets of
intersection points and curves}

On the plane the variety M is defined by two equations

\begin{equation}
\label{C1}
f_{1}(p,q)=0
\end{equation}
and

\begin{equation}
\label{C2} 
f_{2}(p,q)=0
\end{equation}
where the affine coordinates on the plane are denoted by p and q. The
coisotropic deformations of the variety M, i.e. the set of intersection
points of the curves (\ref{C1}) and (\ref{C2}) is defined by the condition

\begin{equation} 
\label{coisoeqnf1f2}
\left\{ f_{1}(p,q;x,y),f_{2}(p,q;x,y)\right\}\mid_{ \Gamma }=0
\end{equation}
where $x$ and $y$ stand for the deformation parameters. The coisotropic
submanifold $\Gamma $ is this case a two-dimensional Lagrangian submanifold.

This construction admits an alternative interpretation. Indeed, let we have
the algebraic curve given by equation (\ref{C1}). One can define deformation of
this curve in the following way. First, we assume that the coefficients of
the polynomial $f_{1}(p,q)$ depend on deformation parameters $x$ and $y$. Then
we take a function $f_{2}(p,q;x,y)$ which is polynomial in p and q. Finally,
we require that the functions $f_{1}$and $f_{2}$ obey the coisotropy
condition (\ref{coisoeqnf1f2}).
\begin{definition}
%\textbf{Definition 2.} 
If the coefficients of the polynomial $f_{1}(p,q)$
are such that the condition (\ref{coisoeqnf1f2}) is satisfied \ then it is said that they
define the coisotropic deformation of the curve (\ref{C1}) generated by the
function $f_{2}(p,q;x,y)$.
\end{definition}
Clearly, this definition is a reciprocal one: if the polynomial $f_{2}$
generates coisotropic deformation of the curve (\ref{C1}), then at the same time
the polynomial $f_{1}$ generates coisotropic deformation of the curve (\ref{C2}).

Let us begin with the simplest case of the second order curves (\ref{C1}) and (\ref{C2}).
As it is well known any nondegenerate quadric is equivalent to
\begin{enumerate}
\item parabola
\begin{equation}
q+p^{2}+ap+b=0
\end{equation}
\item  or to ellipse
\begin{equation}
ap^{2}+bq^{2}+cp+dq+f=0
\end{equation}
\item or to hyperbola
\begin{equation}
\label{hyper}
pq+ap+bq+c=0.
\end{equation}
\end{enumerate}
To construct coisotropic deformation we choose the canonical Poisson bracket
in $R^{4},$ i.e. $\left\{ F,G\right\} =\frac{\partial F}{\partial p}\frac{%
\partial G}{\partial x}+\frac{\partial F}{\partial q}\frac{\partial G}{%
\partial y}-\frac{\partial F}{\partial x}\frac{\partial G}{\partial p}-\frac{%
\partial F}{\partial y}\frac{\partial G}{\partial q}$.

We consider first a parabola and choose $f_{2}$ as an arbitrary polynomial $%
f_{2}(p,q;x,y)$. The submanifold

\begin{eqnarray}
f_{1} =q+p^{2}+a(x,y)p+b(x,y)=0, \\
f_{2}(p,q;x,y) =0
\end{eqnarray}
can be equivalently represented as the zero locus of the functions

\begin{equation}
\label{Par-point}
\begin{split}
f_{1} =q+p^{2}+a(x,y)p+b(x,y)=0, \\
\widetilde{f}_{2} =\sum_{k=0}^{N}\alpha _{k}(x,y)p^{k}=0
\end{split}
\end{equation}
with \ certain N and $\alpha _{k}(x,y)$. In the simplest case N=1, a=0 and $%
\alpha _{1}=1$ the coisotropy condition (\ref{coisoeqnf1f2}) gives $\alpha
_{0x}=0,b_{y}=\alpha _{0y}$, i.e. the deformation is the trivial shift of
the parabola: $b=\beta _{0}(x)+\beta _{1}(y)$ where $\beta _{0}(x)$ and $%
\beta _{1}(y)$ are arbitrary functions.

In the case N=2 one has $\alpha _{2}=1$ and the CS takes the form

\begin{equation}
 \label{eqnparpo}
\begin{split}
\alpha _{1y}-2\alpha _{1}\alpha _{1x}+(a\alpha _{1})_{x}+2(\alpha
_{0}-b)_{x} &=0,  \\
\alpha _{0y}+a\alpha _{0x}-\alpha _{1}b_{x}+2\alpha _{0}(a-\alpha _{1})_{x}
&=0. 
\end{split}
\end{equation}
At $\alpha _{0}=a=b=0$ it is the Burgers-Hopf equation $\alpha _{1y}-2\alpha
_{1}\alpha _{1x}=0$. Equations (\ref{eqnparpo}) describe the coisotropic deformations of
the two points of intersection $(p_{+},q_{+}),(p_{-},q_{-})$  (assuming $%
\alpha _{1}^{2}\geq 4\alpha _{0}$)

\begin{eqnarray}
p_{\pm } &=&-\frac{\alpha _{1}(x,y)}{2}\pm \sqrt{\frac{1}{4}\alpha
_{1}^{2}(x,y)-\alpha _{0}(x,y)},  \notag \\
q_{\pm } &=&(\alpha _{1}-a)p_{\pm }+\alpha _{0}-b
\end{eqnarray}
of the curves (\ref{Par-point}) or equivalently of the parabola

\begin{equation}
q+p^{2}+ap+b=0
\end{equation}
and a straight line

\begin{equation}
q+(a-\alpha _{1})p+b-\alpha _{0}=0.
\end{equation}

Now let us consider the cubic polynomial $f_{2}$, i.e. a set of intersection
points for the curves

\begin{equation}
\label{par}
q+p^{2}+ap+b=0
\end{equation}
and

\begin{equation}
\label{pointcub}
p^{3}+\alpha _{2}p^{2}+\alpha _{1}p+\alpha _{0}=0.
\end{equation}
The CS in this case is of the form

\begin{equation}
\label{parpo3}
\begin{split}
\alpha _{2y}+2\alpha _{1x}-3b_{x}+(a\alpha _{2})_{x}-(\alpha _{2}^{2})_{x}
&=0,   \\
\alpha _{1y}+2\alpha _{0x}+a\alpha _{1x}+2\alpha _{1}a_{x}-2\alpha
_{1}\alpha _{2x}-2\alpha _{2}b_{x} &=0,   \\
\alpha _{0y}+a\alpha _{0x}-\alpha _{1}b_{x}+\alpha _{0}(3a-2\alpha _{2})_{x}
&=0.
\end{split}
\end{equation}

This system contains several reductions of interest. There are two
distinguished between them. The first is given by $a=\alpha _{2}=0$. The
first equation (\ref{parpo3}) then implies that $2\alpha _{1}=3b$ and the rest two
equations take the form

\begin{equation}
\label{KZeqn}
\begin{split}
3b_{y}+4\alpha _{0x} &=0,  \\
4\alpha _{0y}-3(b^{2})_{x} &=0.
\end{split}
\end{equation}
It is the well-known stationary dispersionless Kadomtsev-Petviashvili (KP)
or Khokhlov-Zabolotskaya equation  (see e.g. \cite{KM1,KM2}). \ Equations (\ref{KZeqn})
imply the existence of the function F such that $b=2F_{xx},\alpha _{0}=-%
\frac{3}{2}F_{xy}$ and the system (\ref{KZeqn}) becomes

\begin{equation}
F_{yy}+2(F_{xx})^{2}=0.
\end{equation}
It is the Hirota equation for the stationary dKP equation.

The second reduction is given by the constraints $\alpha _{1}=2b,\alpha _{2}=a$.
The CS (\ref{parpo3}) then is converted to

\begin{equation}
\label{FerMok}
\begin{split}
a_{y}+b_{x} &=0, \\
 b_{y}+\alpha _{0x}&=0, \\
\alpha _{0y}+(a\alpha _{0}-b^{2})_{x} &=0.
\end{split}
\end{equation}
This system of conservation laws implies the existence of the function F
such that

\begin{equation}
a=F_{xxx},\quad b=-F_{xxy},\quad \alpha _{0}=F_{xyy}
\end{equation}
in terms of which it is reduced to the single equation

\begin{equation}
F_{yyy}+F_{xxx}F_{xyy}-(F_{xxy})^{2}=\beta (y)
\end{equation}
where $\beta (y)$ is an arbitrary function. At $\beta =0$ it is the celebrated
WDVV equation \cite{W,DVV,D2}. \ Note that the system (\ref{FerMok}) has appeared for the
first time in the paper \cite{FM}.

Thus, the stationary dKP equation and WDVV\ equation describe coisotropic
deformations of the set of intersection points of the curves (\ref{par}) and (\ref{pointcub}).
\ The locus of common zeros for the polynomials (\ref{par}),(\ref{pointcub}) coincides with
zero locus of the parabola

\begin{equation}
q+p^{2}+ap+b=0
\end{equation}
and the hyperbola

\begin{equation}
\widetilde{f_{2}}=pq+(\alpha _{2}-a)q+(b-\alpha _{1}+a(\alpha
_{2}-a))p+b(\alpha _{2}-a)-\alpha _{0}=0.
\end{equation}
For the stationary dKP case the equation of hyperbola takes the form

\begin{equation}
pq-\frac{1}{2}bp-\alpha _{0}=0
\end{equation}
while at the WDVV\ case one has

\begin{equation}
pq-bp-\alpha _{0}=0.
\end{equation}
It would be of interest to clarify the geometrical difference between the
dKP and WDVV cases.

The above construction shows also that the stationary dKP and WDVV\
equations describe at the same time special classes of coisotropic
deformations for the hyperbola (\ref{hyper}) generated by parabola (\ref{par}).

\ As the last illustrative example in this section we consider coisotropic
deformations of a circle

\begin{equation}
\label{circle}
f_{1}=p^{2}+q^{2}+u=0.
\end{equation}
With the choice

\begin{equation}
\label{2cubic}
f_{2}=p^{3}-3pq^{2}+ap+bq=0
\end{equation}
the CS takes the form

\begin{eqnarray*}
(ua)_{x}+(ub)_{y} &=&0,  \notag \\
3u_{x}-a_{x}+b_{y} &=&0,\\
3u_{y}+a_{y}+b_{x}&=&0.
\end{eqnarray*}
It is the stationary dispersionless Veselov-Novikov (dVN) equation  (see
\cite{KM1}). It describes also the coisotropic deformations of the set of
intersection points of the circle (\ref{circle}) and the cubic (\ref{2cubic}).

\ Considering higher order polynomials $f_{2}$, one gets coisotropic
deformations described by the stationary higher \ dKP and dVN equations.

\section{Deformations of plane cubics}

General form of the plane cubic is ($f_{1}\doteq \zeta $)

\begin{equation}
\label{pcub}
\zeta =p^{2}-q^{3}-u_{4}pq-u_{3}q^{2}-u_{2}p-u_{1}q-u_{0}=0.
\end{equation}
Choosing the linear second equation, i.e. the straight line

\begin{equation}
\label{retta}
f_{2}=\alpha p+\beta q+\gamma =0
\end{equation}
and canonical Poisson bracket, one gets the following CS  ($\alpha =1$)

\begin{equation}
\label{CS31}
\begin{split}
&\beta ^{2}u_{4x}-\beta u_{4y}-2\beta \beta _{y}+3\beta ^{2}\beta
_{x}+3\gamma _{x}-u_{4}\beta _{y}+u_{3}\beta _{x}+\beta u_{3x}-u_{3y}+2\beta
u_{4}\beta _{x}=0, \\
&\beta \gamma u_{4x}-\beta u_{4}\gamma _{x}+3\beta u_{2}\beta
_{x}-u_{1y}+2\gamma u_{4}\beta _{x}-u_{4}\gamma _{y}-u_{2}\beta
_{y}-\gamma
u_{4y}-2u_{3}\gamma _{x}-\beta u_{4y}+2u_{1}\beta _{x}+\\
&+\beta u_{1x}+6\beta \gamma
\beta _{x}-2\gamma \beta _{y}+\beta ^{2}u_{2x}-2\beta \gamma _{y}=0, \\
& \beta u_{0x}-u_{1}\gamma _{x}-u_{2}\gamma _{y}-u_{0y}+3u_{0}\beta
_{x}-\gamma u_{2y}-2\gamma \gamma _{y}+3\gamma ^{2}\gamma
_{y}+\beta \gamma u_{2x}-\gamma u_{4}\gamma _{x}+3\gamma
u_{2}\beta _{x}=0.
\end{split}
\end{equation}

The variety M for the choice (\ref{retta}) consists of at most three points of
intersection of the cubic (\ref{pcub}) with the straight line (\ref{retta}). Coisotropic
deformations of these three points are described by the CS (\ref{CS31}) \ and
generate three surfaces in $R^{4}$.

\bigskip In the particular case $u_{4}=u_{3}=u_{2}=0,\alpha =1,\gamma =0$
the CS system takes the form

\begin{eqnarray}
\beta _{y}-\frac{3}{2}\beta \beta _{x} =0, \\
u_{1y}-\beta u_{1x}-2u_{1}\beta _{x} =0, \\
u_{0y}-\beta u_{0x}-3u_{0}\beta _{x} =0.
\end{eqnarray}

These equations describe deformations of the moduli $u_{1}$and $u_{0}$ of
the elliptic curve. The behavior of the moduli is defined completely by the
solution of the Burgers-Hopf equation for $\beta $. For the discriminant $
\Delta =16(4u_{1}^{3}+27u_{0}^{2})$ of the elliptic curve (see e.g.
\cite{Huse,MKM}) one has the equation

\begin{equation}
\Delta _{y}-\beta \Delta _{x}-6\beta _{x}\Delta =0.
\end{equation}
while deformation of the invariant $j=12^{3}\frac{4u_{1}^{3}}{
4u_{1}^{3}+27u_{0}^{2}}=3^{3}4^{6}\frac{u_{1}^{3}}{\Delta }$ is defined by
the equation

\begin{equation*}
j_{y}-\beta j_{x}=0.
\end{equation*}

We see that at the points of the gradient catastrophe for Burgers-Hopf
equation where $\beta _{x},\beta _{y}\rightarrow \infty $ the moduli $u_{1}$,
 $u_{0}$ \ and the discriminant $\Delta $ exhibit gradient catastrophe behavior too.

Stationary solutions of the CS (\ref{CS31}) with constant $
u_{4},u_{3},u_{2},u_{1},u_{0}$ are of interest too. It would correspond to
Abel's approach to the law of addition on the cubic  (see e.g. \cite{MKM}, section
2.14). In the case $u_{4}=u_{3}=u_{2}=0$ the CS (\ref{CS31}) is reduced to the
system
\begin{eqnarray}
2\beta \beta _{y}+3\beta ^{2}\beta _{x}+3\gamma _{x} &=&0, \nonumber\\
2u_{1}\beta _{x}+6\gamma \beta  \beta _{x}+2(\gamma \beta )_{y} &=&0, \nonumber\\
-u_{1}\gamma _{x}+3u_{0}\beta _{x}+\gamma \gamma _{y}+3\gamma
^{2}\beta _{x} &=&0.
\end{eqnarray}
This overdetermined system, obviously, may have nontrivial solutions only for very
special $\beta ,\gamma $ and constant $u_{0},u_{1}.$

For general polynomial $f_{2}$ the variety M has also the basis

\begin{eqnarray}
\zeta &=&0,  \notag \\
\widetilde{f_{2}} &=&\alpha (q)+\beta (q)p
\end{eqnarray}
where $\alpha (q)$ and $\beta (q)$ are arbitrary polynomials in q.

\section{Deformations of space curves}

In three and more dimensional spaces a zoology of algebraic varieties is
richer and correspondingly their deformations form a much larger collection
of different cases.

In the three-dimensional affine space an algebraic curve is defined by two
polynomial equations

\begin{equation}
f_{1}(p_{1},p_{2},p_{3})=0,\quad f_{2}(p_{1},p_{2},p_{3})=0.
\end{equation}
Coisotropic deformations of this curve are defined by the condition

\begin{equation}
\left\{ f_{1}(p;x),f_{2}(p;x)\right\}\mid_{\Gamma }=0.
\end{equation}
A coisotropic submanifold $\Gamma $ typically is the four-dimensional
submanifold in $R^{6}$.

A simple example is provided by the twisted cubic defined by the
equations

\begin{equation}
\label{twicub1}
f_{1} =p_{2}+p_{1}^{2}+u=0
\end{equation}
and
\begin{equation}
\label{twicub2}
f_{2} =p_{3}+p_{1}^{3}+vp_{1}+w=0
\end{equation}
which is one of the first standard examples in all textbooks on algebraic geometry. 
\ The CS with the choice of
canonical Poisson bracket in $R^{6}$ in this case is given by the equations

\begin{equation}
\label{twisted}
\begin{split}
u_{x_{3}}+vu_{x_{1}}u-w_{x_{2}} &=0, \\
v_{x_{2}}+2w_{x_{1}} &=0, \\
3u_{x_{2}}-2v_{x_{1}} &=0.
\end{split}
\end{equation}
So, $v=\frac{3}{2}u$ and one has the system

\begin{equation}
\begin{split}
u_{x_{3}}+\frac{3}{2}u_{x_{1}}u-v_{x_{2}} &=0, \\
3u_{x_{2}}+4v_{x_{1}} &=0
\end{split}
\end{equation}
which is the dKP equation. Thus, the three-dimensional coisotropic
deformations of the twisted cubic in $R^{3}$ are doverned by the dKP
equation. For polynomial solutions of the dKP equation the family of the deformed varieties in $R^{6}$ defined by the equations

\begin{eqnarray}
f_{1} &=&p_{2}+p_{1}^{2}+u(x_{1},x_{2},x_{3})=0, \\
f_{2} &=&p_{3}+p_{1}^{3}+\frac{3}{2}u(x_{1},x_{2},x_{3})p_{1}+w(x_{1},x_{2},x_{3})=0
\end{eqnarray}
(i.e. the submanifold $\Gamma$) is the algebraic variety too.

 We note that on the plane $p_2,p_3$ the twisted cubic (\ref{twicub1}),(\ref{twicub2}) is the cubic curve given by the equation
 
\begin{equation*}
{p_{{3}}}^{2}+{p_{{2}}}^{3}+ \left( 3\,u-2\,v \right) {p_{{2}}}^{2}+2
\,wp_{{3}}+ \left( {v}^{2}+3\,{u}^{2}-4\,vu \right) p_{{2}}+({u}^{3}+{w
}^{2}-2\,v{u}^{2}+{v}^{2})=0.
\end{equation*}
It is obvious, however, that this cubic curve is degenerate for all the values of $u$ and $v$ since it has polynomial parametrization given by (\ref{twicub1}), (\ref{twicub2}). Deformations of nondegenerate elliptic curves  will be studied in next section.

Choosing

\begin{equation}
f_{2}=p_{n}+p_{1}^{n}+\sum_{k=0}^{n-2}v_{k}p_{1}^{k},
\end{equation}
one constructs the coisotropic deformations of the n-th order curve in $%
R^{3} $. These deformations are described by the higher dKP equations.

\ Another simple example corresponds to

\begin{eqnarray}
f_{1} &=&p_{1}^{2}+p_{2}^{2}+u=0, \\
f_{2} &=&p_{3}+p_{1}^{3}-3p_{1}p_{2}^{2}+ap_{1}+bp_{2}=0.
\end{eqnarray}
This curve M is the intersection of the cylinder defined by the first
equation and cubic surface given by the second equation.

The coisotropy condition gives rise to the following CS

\begin{eqnarray*}
u_{x_{3}}+(ua)_{x_{x}}+(ub)_{x_{2}} &=&0, \\
3u_{x_{1}}-a_{x_{1}}+b_{x_{2}} &=&0,\\
 3u_{x_{2}}+a_{x_{2}}+b_{x_{1}}&=&0.
\end{eqnarray*}
It is the dVN equation  (see e.g. \cite{KM2}). For higher order $f_{2}$
coisotropic deformations are described by the higher dVN equations.

Finally, let us consider the case

\begin{eqnarray*}
f_{1} &=&p_{1}p_{2}+up_{1}+v=0, \\
f_{2} &=&p_{3}+\alpha p_{1}^{2}+\beta p_{2}^{2}+ap_{1}+b=0.
\end{eqnarray*}
This curve is the intersection of cylindric hyperbola and paraboloid. Its
coisotropic deformations is described by the following CS

\begin{eqnarray*}
u_{x_{3}}+au_{x_{1}}-\beta (u^{2})_{x_{2}}+2\alpha v_{x_{1}}-b_{x_{2}} &=&0,
\\
v_{x_{3}}+(av)_{x_{1}}-2\beta (uv)_{x_{2}} &=&0, \\
a_{x_{2}}-2\alpha u_{x_{1}} &=&0, \quad b_{x_{1}}-2\beta v_{x_{2}}=0.
\end{eqnarray*}
This hydrodynamical system has distinguished reductions. At $a=0,\alpha
=0,\beta =\frac{1}{2}$ it is the 2+1-dimensional generalization of the
one-layer Benney system proposed in \cite{K2,Z}. At $\beta =-\alpha =\frac{1}{2}$
it is the dispersionless Davey-Stewartson system considered in \cite{Kono}.

 We note that deformations considered in this section can be treated also as coisotropic deformations
of a surface given by equation $f_{1}(p_{1},p_{2},p_{3})=0$ generated by a surface defined by equation
$f_{2}(p_{1},p_{2},p_{3})=0$ (or vice versa).

\section{Coisotropic deformations of elliptic curve. Two-dimensional case}

Now we will consider coisotropic deformations of nondegenerate cubics. This case is of importance
since it provides us with the example of deformations for algebraic curve of nonzero genus.
 
 General cubic is given by equation (\ref{pcub}), i.e.
 
\begin{equation}
\label{ell-P}
\begin{split}
\mathcal{E}&={p_3}^2-{p_2}^3-u_4 p_2 p_3-u_3 {p_2}^2-u_2 p_3-u_1 p_2-u_0. 
\end{split}
\end{equation}
where for further convenience we have changed the notation ($p=p_3,q=p_2$).
 To construct coisotropic deformations of the cubic (\ref{ell-P}) we choose the canonical Poisson structure in $R^{6}$
and functions

\begin{equation}
\label{gav}
f_n=p_n-\alpha_{n}(p_2)-\beta_{n}(p_2)p_3=0
\end{equation}
where $\alpha$ and $\beta$ are polynomials in $p_2$ with coefficients depending on the deformation variables $x_2,x_3,x_n$.

 It is quite instructive to consider first the particular deformations which correspond to a cyclic variable $x_2$, i.e. when $u_i=u_i(x_3,x_n)$. In this case $p_2$ appears in (\ref{ell-P}), (\ref{gav}) as a parameter $\lambda=p_2$ and the 
corresponding deformations of the cubic (\ref{ell-P}) are two-dimensional.

So we consider the cubic
\begin{equation}
\mathcal{E}=p_3^{2}-\lambda^{3}-u_4\lambda p_3-u_3\lambda^{2}-u_2p_3-u_1\lambda-u_0=0.
\end{equation}
We choose the generating function (\ref{gav}) as

\begin{equation}
 f_{5}=p_5-v_{3}\lambda^{2}-v_1\lambda-v_0-(v_2+\lambda)p_3
\end{equation}
in order to get a CS allowing deformations for all coefficients $u_i$ of the cubic. 
 The coisotropy condition
$$
\{f_{5},\mathcal{E}\}|_{\Gamma}=0
$$
gives the system 
\begin{equation}
\label{1+1gendef}
 \begin{split}
& {\frac {\partial {u_{{4}}}}{\partial x_3}}  
+2\,{\frac {\partial v_{{3}}}{\partial x_3}}  =0 \\ 
& {\frac {\partial u_{{3}}}{\partial x_3}}   
+2\,{\frac {\partial v_{{2}}}{\partial x_3}}  -u_{{4}}   {\frac {\partial v_{{3}}}{
\partial x_3}}  =0\\ 
& u_{{4}}   {\frac {\partial v_{{2}}}{\partial x_3}}
   -{\frac {\partial u_{{4}}}{\partial x_5}}   
+   {\frac {\partial u_{{4}}}{\partial x_{{3}}}}      v_{{2}}  +{\frac {\partial u_{{2}}}{\partial x_3}}  +2\,{\frac {\partial v_{{1}}}{\partial x_3}} =0\\
& -{\frac {\partial u_{{3}}}{\partial x_5}}   
+{\frac {\partial u_{{1}}}{\partial x_3}}   -
u_{{2}}   {\frac {\partial v_{{3}}}{\partial x_3}}
   +   {\frac {\partial u_{{3}}}{\partial 
x_3}}      v_{{2}}  -u_{{4}}   {\frac {\partial v_{{1}}}{\partial x_3}}   
 +2\,   {\frac {\partial v_{{2}}}{\partial x_3}}   
   u_{{3}}   =0\\
& -{\frac {\partial u_{{2}}}{\partial x_5}}   +2\,{\frac {\partial v_{{0}}}{\partial x_3}}   +
   {\frac {\partial u_{{2}}}{\partial x_3}}     v_{{2}}   
+u_{{2}} {\frac {\partial v_{{2}}}{\partial x_3}} =0\\
& -{\frac {\partial u_{{1}}}{\partial x_5}}   +{\frac {\partial u_{{0}}}{\partial x_3}}   +
   {\frac {\partial u_{{1}}}{\partial x_3}}  v_{{2}}   
-u_{{4}}    {\frac {\partial v_{{0}}}{\partial x_3}}   
-u_{{2}}   {\frac {\partial v_{{1}}}{\partial x_3}}    
+2\,   {\frac {\partial v_{{2}}}{\partial x_3}}   u_{{1}}   =0\\
&-{\frac {\partial u_{{0}}}{\partial x_5}}   +
   {\frac {\partial u_{{0}}}{\partial x_3}}   v_{{2}}   -u_{{2}} {\frac {\partial v_{{0}}}{\partial x_3}}  +2\,   {\frac {\partial v_2}{\partial x_3}}   u_{{0}}  =0.
 \end{split}
\end{equation}
In this and next sections in order to avoid  triple indices we write the derivatives in the explicit way.
The first two equations imply that  

\begin{equation*}
 v_3=-\frac{1}{2}u_4 \qquad v_2=-\frac{1}{2}u_3+\frac{1}{4}{u_4}^2. 
\end{equation*}
and the  system (\ref{1+1gendef}) becomes the system of five equations for $u_0,u_1,u_2,u_3,u_4$. The fields $v_1$ and $v_0$ can be considered as a couple of gauge fields.

 This system admits several reductions. The most interesting one  corresponds to the constraint $u_4=0,u_2=0$. In this case  $v_3=0,\ 2v_2=-u_3,\ 2v_1=-u_2,\ v_0=0$ and the above system is converted into the following

\begin{equation}
\label{3KdV}
 \begin{split}
 {\frac {\partial u_{{3}}}{\partial x_5}}   &= 
-\frac{3}{2}\,   {\frac {\partial u_{{3}}}{\partial x_3}}u_{{3}} +{\frac {\partial u_{{1}}}{\partial x_3}},    \\
{\frac {\partial u_{{1}}}{\partial x_5}}   &=
-\frac{1}{2}\,   {\frac {\partial u_{{1}}}{\partial x_3}}       
u_{{3}}  -   {\frac {\partial u_{{3}}}{\partial x_3}}    
   u_{{1}}     +{\frac {\partial u_{{0}}}{\partial x_3}},     \\
{\frac {\partial u_{{0}}}{\partial x_5}}   &=
-   {\frac {\partial u_{{3}}}{\partial x_3}}u_{{0}} -\frac{1}{2}\,   {\frac {\partial u_{{0}}}{\partial x_3}} u_{{3}}   
 \end{split}
\end{equation}
which is the well known 3 component dispersionless KdV equation  (see e.g.(\cite{FP}). Solutions of this system describe  1+1 dimensional coisotropic deformations of the elliptic curve $\mathcal{E}_r={p_3}^2-\lambda^{3}-u_3 \lambda^{2}-u_1 \lambda-u_0. $
generated by the standard symplectic form and the polynomial ${f_{5}}_r=p_5-\lambda p_{3}+\frac{u_3}{2}p_3$.
 \\
This evolution coincides with that  obtained earlier in a totally different manner in \cite{KK}. In order to complete the comparison with the results presented in the paper \cite{KK} we will show how coisotropic deformations give rise to corresponding system in terms of Riemann invariants. First we present the elliptic curve in the form $\mathcal{E}_{e}={p_3}^2-(\lambda-e_1)(\lambda-e_2)(\lambda-e_3)$. Then it is a simple check that the coisotropy condition
\begin{equation}
 \{ \mathcal{E}_{e} , f_{5}\}|_{\Gamma} = 0
\end{equation}
is equivalent to the 3 component dKP system in terms of Riemann invariants, i.e

\begin{equation}
\label{1+1De1e2e3}
\begin{split}
 \frac{\partial e_1}{\partial x_5}  &= \left( \frac{3}{2}e_1 + \frac{1}{2}e_2 + \frac{1}{2}e_3\right) 
\frac{\partial e_1}{\partial x_3}, \\
 \frac{\partial e_2}{\partial x_5}  &= \left( \frac{1}{2}e_1 + \frac{3}{2}e_2 + \frac{1}{2}e_3\right) 
 \frac{\partial e_2}{\partial x_3},  \\
\frac{\partial e_3}{\partial x_5}  &= \left( \frac{1}{2}e_1 + \frac{1}{2}e_2 + \frac{3}{2}e_3\right) 
\frac{\partial e_3}{\partial x_3} . 
\end{split}
\end{equation}

 From the algebro-geometrical characterization viewpoint three of five parameters $u_i, i=0,1,2,3,4$ for an elliptic curve are redundant (see e.g.\cite{Huse}). Only two special combinations of $u_i$ (moduli $g_2$ and $g_3$) are essential and the canonical form of elliptic curve is  

\begin{equation}
\label{canform}
 \mathcal{E}_c={\pi_3}^2-({\pi_2}^3+ g_2 \pi_2 + g_3)=0.
\end{equation}
 Moduli $g_2$ and $g_3$ are given by the formulae \cite{Huse}
\begin{equation}
  g_2= u_1 -\frac{1}{3}{u_3}^2 \qquad g_3= u_0+\frac{2}{27} {u_3}^3 -\frac{1}{3}u_1u_3.
\end{equation}
where for sake of simplicity we choose $u_4=u_2=0$.
General cubic (\ref{ell-P}) is converted into the canonical form by admissible transformation ($u_4=u_2=0$) \cite{Huse}

\begin{equation}
\label{Moebius}
  p_2= \pi_2-\frac{1}{3}u_3 \quad p_3=\pi_3. 
\end{equation}

Direct calculation gives the following equations for moduli

\begin{equation}
\label{1+1Dg2g3}
 \begin{split}
  \frac{\partial g_2}{\partial x_5}&= {\frac {\partial g_{{3}}}{\partial x_3}}
 -\frac{5}{6}\, {\frac {\partial g_{{2}}}{\partial x_3}} u_{{3}}\
 -\frac{2}{3}\,   {\frac {\partial u_{{3}}}{\partial x_3}}   g_{{2}},\\
  \frac{\partial g_3}{\partial x_5}&= -\frac{5}{6}\, {\frac {\partial g_{{3}}}{\partial x_3}}u_{{3}} -\frac{1}{3}
  {\frac{\partial g_{{2}}}{\partial x_3}} g_{{2}} -   {\frac {\partial g_{{3}}}{\partial x_3}}u_{{3}},    \\
  \frac{\partial u_3}{\partial x_5}&={\frac {\partial g_2}{\partial x_3}}  -\frac{5}{6}\,   {\frac {\partial u_3}{
\partial x_3}}     u_3. 
 \end{split}
\end{equation}
This system contains not only moduli but also the function $u_3$. One can, in principal, express $u_3$ in terms of $g_2$ and $g_3$ solving the third of above equations  (Burgers-Hopf equation with the source) or equivalently solving the Hamilton-Jacobi equation 
\begin{equation}
\phi_{x_{5}}+\frac{5}{12}(\phi_{x_{3}})^2-g_2=0
\end{equation}
where $\phi_{x_{3}}=u_3$.
  
 Thus, the system (\ref{1+1Dg2g3}) governs the coisotropic deformations of moduli of elliptic curve parametrized
by two variables $x_3$ and $x_5$. Equations (\ref{1+1Dg2g3}) allows us also to find deformations of the discriminant $\delta$ and invariant J of elliptic curve. The corresponding equations are rather complicated and we do not present them here. For illustration we consider the following simple solution of the system (\ref{1+1Dg2g3}) (C is an arbitrary constant)

\begin{equation}
 \begin{split}
  u_3&=C,\\
  u_1&=x_5-\frac{C^2}{4},\\
  u_0&=x_3-\frac{C}{2}x_5+\frac{C^3}{6} 
 \end{split}
\end{equation}
which provides us with simple, linear deformation of the elliptic curve.
For this solution the discriminant is equal to 
\begin{equation}
 \Delta(x_3,x_5)=16(4\,{x_5}^{3}+{\frac {47}{4}}\,{x_5}^{2}{C}^{2}-\frac{21}{2}\,x_5{C
}^{4}+{\frac {49}{24}}\,{C}^{6}+27\,{x_3}^{2}-45\,x_3x_5C+
{\frac {35}{2}}\,x_3{C}^{3}).
\end{equation}
In the particular case $C=0$ we have $\Delta(0,0) = 0$ and so at $x_3=x_5=0$ the curve is singular. Deformation
 (variation of $x_3$ and $x_5$) however generically desingularize it.
Conversely, if $C \neq 0$ then $\Delta(0,0) \neq 0$, but the deformation produces $\Delta$ which may be different from zero not everywhere. This deformation changes $J$ and, therefore, changes the elliptic curve. We note also that for this solution the family of deformed cubics is the algebraic variety too.

 Thus, it is very natural to look for coisotropic deformations of moduli starting directly from the canonical form (\ref{canform}) of elliptic curve. One immediately finds that the canonical Poisson structure considered before is not appropriate in this case since the coisotropy condition with such a Poisson bracket gives rise to only trivial deformations. Thus, one should search for an adequate Poisson structure. 

 One way to find it is to study transformation of the canonical Poisson structure under the transformations of Darboux coordinates $p_i,x_i$ to new coordinates $\pi_i,\tau_i$ given by the formula  (\ref{Moebius}) and by $\pi_5=p_5$, $\tau_i=x_i,\ i=2,3,5$. Direct calculation gives the following transformed Poisson structure

\begin{equation}
 \begin{split}
\label{taupi}
  \{ \tau_i, \tau_j\}_{\tau\pi} &=0, \qquad i,j=2,3,5\\
  \{ \tau_i, \pi_j\}_{\tau\pi} &=\delta _{ij}, \qquad i,j=2,3,5 \\ 
  \{ \pi_i, \pi_j\}_{\tau\pi} &=\frac{1}{3}\frac{\partial u_3}{\partial x_3} \delta_{i3}\delta_{j2}
+\frac{1}{3}\frac{\partial u_3}{\partial x_5} \delta_{i5}\delta_{j2}, \qquad i,j=2,3,5.
 \end{split}
\end{equation}
 It is not difficult to check that the coisotropy condition

\begin{equation}
 \{ \mathcal{E}_c,J^{(5)} \}_{ \tau \pi }|_{\Gamma}=0
\end{equation}
with the Poisson bracket (\ref{taupi}) and $J^{(5)}|_{p_2=\pi_2-\frac{1}{3}u_3}=\pi_5-\pi_{2}\pi_{3}+\frac{u_3}{6}\pi_3$ gives the system (\ref{1+1Dg2g3}).

 One can perform similar transformations in other cases. The problem of consistency of an algebraic curve and corresponding Poisson structure which allow to construct nontrivial coisotropic deformations will be discussed elsewhere.

 Finally, we note that the results of this section can be extended to hyperelliptic curves. Indeed, of one takes a hyperelliptic curve 

\begin{equation}
 {p_{2n+1}}^2={p_2}^{2n+1}+\sum_{i=0}^{2n} v_i(x_{2n+1},x_{2n+3}){p_2}^i,
\end{equation}
choose the  polynomial functions of the form  $f_m =p_{2n+3}-(\sum_{i=0}^{m} u_i(x_{2n+1},x_{2n+3}){p_2}^i) p_{2n+1}$ then the coisotropy condition with the canonical Poisson bracket reproduces hydrodynamical type systems derived by a different method in the paper \cite{KK}.  

\section{Three-dimensional deformations of elliptic curves and Poisson structures}

We will consider now fully three-dimensional deformations of the general cubic (\ref{ell-P}). We take the canonical Poisson bracket in $R^6$, choose the generating function as $f_4=p_4-{p_2}^2-u_2p_3-u_1p_2-u_0$ and denote deformation variables by $x_2,x_3,x_4$. The coisotropy condition takes the form

 \begin{equation}  
\label{eqnKM}
 \{ p_4-P_4(p_2,p_3), \mathcal{E} \}|_\Gamma = c_7\, {p_2}^2p_3 + c_6\, {p_2}^3 + c_5\, p_2 p_3+c_4\, {p_2}^2 +c_3\, p_3+c_2\, p_2+c_0=0
\end{equation}

where
\begin{equation}  
\label{eqnKM-ci}
\begin{split}
c_7=&2\,{\frac {\partial u_{{4}}}{\partial x_2}} -3\,{\frac {\partial v_{{2}}}{\partial x_2}} \\
c_6=&  2\,{\frac {\partial u_{{3}}}{\partial x_2}}  -3\,{\frac {\partial v_{{1}}}{\partial x_2}}
  -u_{{4}}  {\frac {\partial v_{{2}}}{\partial x_2}}  +2\,{\frac {\partial v_{{2}}}{\partial x_{{3}}}}  \\
c_5 =& v_{{1}}  {\frac {\partial u_{{4}}}{\partial x_2}}   
 -u_{{4}}  {\frac {\partial v_{{1}}}{\partial x_2}}  
-2\,u_{{3}}  {\frac {\partial v_{{2}}}{\partial x_2}}  
+v_{{2}}  {\frac {\partial u_{{4}}}{\partial x_3}}  
+u_{{4}}  {\frac {\partial v_{{2}}}{\partial x_3}}  - {u_4}^2{\frac {\partial v_{{2}}}{\partial x_2}}  -{\frac {\partial u_{{4}}}{\partial x_4}}  +2\,{\frac {\partial u_{{2}}}{\partial x_2}}  
+2\,{\frac {\partial v_{{1}}}{\partial x_3}} \\
c_4=& -u_{{4}}   u_{{3}} {\frac {\partial v_{{2}}}{\partial x_2}} 
+v_{{1}}  {\frac {\partial u_{{3}}}{\partial x_2}}  
-{\frac {\partial u_{{3}}}{\partial x_4}} +2\,{\frac {\partial u_{{1}}}{\partial x_2}} 
-3\,{\frac {\partial v_{{0}}}{\partial x_2}}  
-2\,u_{{3}}  {\frac {\partial v_{{1}}}{\partial x_2}}  
+v_{{2}}  {\frac {\partial u_{{3}}}{\partial x_3}} -u_{{4}}  {\frac {\partial v_{{1}}}{\partial x_3}}  
+2\,u_{{3}} {\frac {\partial v_{{2}}}{\partial x_3}} \\
c_3=&-u_{{4}} u_{{2}} {\frac {\partial v_{{2}}}{\partial x_2}}
 -{\frac {\partial u_{{2}}}{\partial x_4}}  +2\,{\frac {\partial v_{{0}}}{\partial x_3}} 
+ v_{{1}} {\frac {\partial u_{{2}}}{\partial x_2}}
  -u_{{4}}  {\frac {\partial v_{{0}}}{\partial x_2}} -u_{{1}}  {\frac {\partial v_{{2}}}{\partial x_2}}
 +v_{{2}}  {\frac {\partial u_{{2}}}{\partial x_3}}  +u_{{2}}  {\frac {\partial v_{{2}}}{\partial x_3}} \\
c_2=& -u_{{4}}  u_{{1}} {\frac {\partial v_{{2}}}{\partial x_2}} 
 -{\frac {\partial u_{{1}}}{\partial x_4}} 
+2\,{\frac {\partial u_{{0}}}{\partial x_2}} 
 + v_{{1}} {\frac {\partial u_{{1}}}{\partial x_2}}
 -2\,u_{{3}}  {\frac {\partial v_{{0}}}{\partial x_2}}
  -u_{{1}}  {\frac {\partial v_{{1}}}{\partial x_2}}
+v_{{2}}  {\frac {\partial u_{{1}}}{\partial x_3}} -u_{{4}} {\frac {\partial v_{{0}}}{\partial x_3}} 
-u_{{2}}  {\frac {\partial v_{{1}}}{\partial x_3}} 
+2\, u_{{1}} {\frac {\partial v_{{2}}}{\partial x_3}} \\
c_0=& v_{{1}} {\frac {\partial u_{{0}}}{\partial x_2}}
-u_{{1}}  {\frac {\partial v_{{0}}}{\partial x_2}}  
+v_{{2}}  {\frac {\partial u_{{0}}}{\partial x_3}}  
-u_{{2}}  {\frac {\partial v_{{0}}}{\partial x_3}} 
-{\frac {\partial u_{{0}}}{\partial x_4}} 
-u_{{4}} u_{{0}} {\frac {\partial v_{{2}}}{\partial x_2}}  
+2\, u_{{0}} {\frac {\partial v_{{2}}}{\partial x_3}} .
\end{split}
\end{equation}

Thus, the coisotropy  condition (\ref{eqnKM}) is equivalent to  the equations $c_i=0$. The conditions $c_7=0$ and $c_6=0$ imply that
\begin{equation}
 v_2=\frac{2}{3} u_4 \quad , \qquad v_1=\frac{2}{3}u_3-\frac{2}{9}{u_4}^2+\frac{4}{3}{\frac{\partial}{\partial x_2}}^{-1}\frac{\partial u_4}{\partial x_3}.
\end{equation}
The rest of these conditions give rise to the system
\begin{equation}
\label{KP-1}
 \begin{split}
 {\frac {\partial u_{{4}}}{\partial x_4}} =& -\frac{2}{3}\frac{\partial}{\partial x_2}(u_2u_3) -\frac{5}{9}{u_4}^2 \frac{\partial u_4}{\partial x_2} 
+\frac{4}{9} u_4 \frac{\partial u_4}{\partial x_3}  +2\frac{\partial u_2}{\partial x_2}  +\frac{4}{3} \frac{\partial u_3}{\partial x_3} +\frac{4}{9}\frac{\partial u_4}{\partial x_2}   {\frac{\partial}{\partial x_2}}^{-1} \frac{\partial u_4}{\partial x_3} +\frac{8}{9} {\frac{\partial}{\partial x_2}}^{-1} {\frac{\partial^2 u_4}{\partial {x_3}^2}}  \\ 
%%%%%%%%%%%%%%%%%%%%%%%%%%%%%%%%%%%%%%%%%%%%%%%%%%%%%%%%%%%%%%%%%%%%%%%%%%%%%%%%%%%%%%%%%%%%%%%%%%%%%%%%%%%
 {\frac {\partial u_{{3}}}{\partial x_4}} =& -\frac{2}{3} u_{{4}}   u_{{3}} {\frac {\partial u_{{4}}}{\partial x_2}} +v_{{1}}  {\frac {\partial u_{{3}}}{\partial x_2}}+2\,{\frac {\partial u_{{1}}}{\partial x_2}} 
-3\,{\frac {\partial v_{{0}}}{\partial x_2}}  -2\,u_{{3}}  {\frac {\partial v_{{1}}}{\partial x_2}}  
+\frac{2}{3} u_4  {\frac {\partial u_{{3}}}{\partial x_3}} 
-u_{{4}}  {\frac {\partial v_{{1}}}{\partial x_3}}  
+\frac{4}{3} \,u_{{3}} {\frac {\partial u_4}{\partial x_3}}  \\
 %%%%%%%%%%%%%%%%%%%%%%%%%%%%%%%%%%%%%%%%%%%%%%%%%%%%%%%%%%%%%%%%%%%%%%%%%%%%%%%%%%
{\frac {\partial u_{{2}}}{\partial x_4}} =& -\frac{2}{3}u_{{4}} u_{{2}} {\frac {\partial u_{{4}}}{\partial x_2}} 
+2\,{\frac {\partial v_{{0}}}{\partial x_3}} + v_{{1}} {\frac {\partial u_{{2}}}{\partial x_2}}
  -u_{{4}}  {\frac {\partial v_{{0}}}{\partial x_2}} -\frac{2}{3}u_{{1}}  {\frac {\partial u_{{4}}}{\partial x_2}}
 +\frac{2}{3}u_{{4}}  {\frac {\partial u_{{2}}}{\partial x_3}}  +\frac{2}{3}u_{{2}}  {\frac {\partial u_{{4}}}{\partial x_3}} \\ 
%%%%%%%%%%%%%%%%%%%%%%%%%%%%%%%%%%%%%%%%%%%%%%%%%%%%%%%%%%%%%%%%%%%%%%%%%%%%%%%%%%%%%%%%%
 {\frac {\partial u_{{1}}}{\partial x_4}} =& -\frac{2}{3}u_{{4}}  u_{{1}} {\frac {\partial u_{{4}}}{\partial x_2}} 
+2\,{\frac {\partial u_{{0}}}{\partial x_2}} 
 + v_{{1}} {\frac {\partial u_{{1}}}{\partial x_2}}
 -2\,u_{{3}}  {\frac {\partial v_{{0}}}{\partial x_2}}
  -u_{{1}}  {\frac {\partial v_{{1}}}{\partial x_2}}
+\frac{2}{3}u_{{4}}  {\frac {\partial u_{{1}}}{\partial x_3}} 
-u_{{4}} {\frac {\partial v_{{0}}}{\partial x_3}} 
-u_{{2}}  {\frac {\partial v_{{1}}}{\partial x_3}} 
+\frac{4}{3}\, u_{{1}} {\frac {\partial u_{{4}}}{\partial x_3}} \\ 
%%%%%%%%%%%%%%%%%%%%%%%%%%%%%%%%%%%%%%%%%%%%%%%%%%%%%%%%%%%%%%%%%%%%%%%%%%%%%%%%%%%%%
 {\frac {\partial u_{{0}}}{\partial x_4}} =& v_{{1}} {\frac {\partial u_{{0}}}{\partial x_2}}
-u_{{1}}  {\frac {\partial v_{{0}}}{\partial x_2}}  
+\frac{2}{3}u_{{4}}  {\frac {\partial u_{{0}}}{\partial x_3}}  
-u_{{2}}  {\frac {\partial v_{{0}}}{\partial x_3}}
-\frac{2}{3}u_{{4}} u_{{0}} {\frac {\partial u_{{4}}}{\partial x_2}}  
+\frac{4}{3}\, u_{{0}} {\frac {\partial u_{{4}}}{\partial x_3}} 
\end{split}
\end{equation}
where again $v_1=\frac{2}{3}u_3-\frac{2}{9}{u_4}^2+\frac{4}{3}{\frac{\partial}{\partial x_2}}^{-1}\frac{\partial u_4}{\partial x_3}$.\\
This system governs three-dimensional coisotropic deformations of the elliptic curve (\ref{ell-P}). From the viewpoint of deformations of space curves it describes deformations of a space curve which is the intersection of a cylindrical surface generated by the cubic (\ref{ell-P}) and a space quadric defined by the equation $f_4=p_4-{p_2}^2-u_2p_3-u_1p_2-u_0=0$.

 We note  that this system contains an arbitrary field $v_0$ . It is associated with the  gauge freedom for the system. In the gauge 
\begin{equation}
 v_0=\frac{1}{27}u_4^2u_3-\frac{1}{6}u_3^2-\frac{2}{3}u_1-\frac{4}{9}{\frac{\partial}{\partial x_2}}^{-1} \frac{\partial}{\partial x_3}\left(u_2-\frac{1}{27}u_4^3+\frac{1}{6}u_3u_4 \right)
\end{equation}
this system coincides with the dispersionless limit of the first hidden KP system considered in \cite{KMAM}. So,  it is natural to refer to the system  (\ref{KP-1}) as the genus 1 KP system. 

 Now we will extend the observation on the transformation of the Poisson structure done in the previous section to the three-dimensional case and will analyze relation between different Poisson structures which produce the same integrable systems.
 We consider the general cubic (\ref{ell-P}). As before admissible transformations are those which do not change the genus. Let us denote new coordinates by $\xi_{2,3,4},\pi_{2,3,4}$. The only admissible transformations are the
graded transformations $\pi_2=p_2+\alpha(x_2,x_3,x_4)$ and
$\pi_3=p_3+\beta_1(x_2,x_3,x_4)p_2+\beta_0(x_2,x_3,x_4)$. 
In order to preserve the gradation we perform similar change of the coordinate $\pi_4=p_4 + 
\gamma_3(x_2,x_3,x_4){p_2}^2+\gamma_2(x_2,x_3,x_4)p_3 +\gamma_1(x_2,x_3,x_4) p_2+\gamma_0(x_2,x_3,x_4)$. Finally, the polynomiality in the $\pi_i$s variables of the family of Poisson tensors is
preserved only if an adequate change of variables $\xi_i=\xi_i(x_2,x_3,x_4)$ for $i=2,3,4$ is performed.  \\
The inverse to such transformation is of the form
\begin{equation}
 \begin{split}
  x_i&=x_i(\xi_2,\xi_3,\xi_4) \qquad i=2,3,4 \\
  p_2&=\pi_2-\alpha \\
  p_3&=\pi_3-\beta_1 \pi_2 +(\alpha\beta_1-\beta_0)\\
  p_4&=\pi_4-\gamma_3 {\pi_2}^2-\gamma_2 \pi_3 +\left( 2\alpha \gamma_3 +\gamma_2 \beta_1-\gamma_1\right) \pi_2
+\left( -\alpha^2 \gamma_3-\alpha \beta_1 \gamma_2 +\beta_0 \gamma_2+\alpha \gamma_1 -\gamma_0 \right)
 \end{split}
\end{equation}
where $\alpha=\alpha(x=x(\xi))$ and analogously for the other coefficients.
Under this transformation the  elliptic curve is converted to
\begin{equation}
\begin{split}
\label{OO}
 \mathcal{E}_g=& {\pi_3}^2-{\pi_2}^3-(u_{{4}}+2\,\beta_{{1}})\pi_2\pi_3-(-u_{{4}}\beta_{{1}}-{\beta_{{1}}}^{2}-3\,\alpha+u_{{3}}){\pi_2}^2-(-u_{{4}}\alpha+u_{{2}}-2\,\beta_{{1}}\alpha+2\,\beta_{{0}})\pi_3 \\
&-(-2\,\beta_{{1}}\beta_{{0}}-u_{{4}}\beta_{{0}}+2\,u_{{4}}\beta_{{1}}
\alpha+2\,{\beta_{{1}}}^{2}\alpha+3\,{\alpha}^{2}+u_{{1}}-2\,u_{{3}}
\alpha-u_{{2}}\beta_{{1}})\pi_2\\
&-(u_{{2}}\beta_{{1}}\alpha+u_{{0}}+2\,\beta_{{1}}\alpha\,\beta_{{0}}-{
\beta_{{1}}}^{2}{\alpha}^{2}-{\beta_{{0}}}^{2}+u_{{3}}{\alpha}^{2}-u_{
{2}}\beta_{{0}}-{\alpha}^{3}-u_{{4}}\beta_{{1}}{\alpha}^{2}+u_{{4}}
\alpha\,\beta_{{0}}-u_{{1}}\alpha)
\end{split}
\end{equation}
and the deformation function becomes
\begin{equation*}
\begin{split}
 {J^{(4)}}_g=& \pi^4-(\gamma_{{3}}+1){\pi_2}^2-(\gamma_{{2}}+v_{{2}})\pi_3-(-2\,\gamma_{{3}}\alpha+\gamma_{{1}}
-\gamma_{{2}}\beta_{{1}}-2\,\alpha-v_{{2}}\beta_{{1}}+v_{{1}})\pi_2\\
&-(\gamma_{{3}}{\alpha}^{2}+\gamma_{{2}}\beta_{{1}}\alpha-\gamma_{{2}}
\beta_{{0}}-\gamma_{{1}}\alpha+\gamma_{{0}}+{\alpha}^{2}+v_{{2}}\beta_
{{1}}\alpha-v_{{2}}\beta_{{0}}-v_{{1}}\alpha+v_{{0}}).
\end{split}
\end{equation*}
Then the canonical Poisson structure is transformed into the following
{\small
\begin{equation}
\label{001}
 \begin{split}
 \{\xi_i,\xi_j\}=&0 \qquad i,j=2,3,4 \\
 \{\xi_i,\pi_2\}=& \frac{\partial \xi_i}{\partial x_2}\Bigg{|}_{x=x(\xi)} \qquad i=2,3,4\\
\{\xi_i,\pi_3\}=& \frac{\partial \xi_i}{\partial x_3}\Bigg{|}_{x=x(\xi)} +
                  \beta_1 \frac{\partial \xi_i}{\partial x_2}\Bigg{|}_{x=x(\xi)}\qquad i=2,3,4\\
\{\xi_i,\pi_4\}=& 2\gamma_3\frac{\partial \xi_i}{\partial x_2}\Bigg{|}_{x=x(\xi)} \pi_2 + 
\frac{\partial \xi_i}{\partial x_4}\Bigg{|}_{x=x(\xi)}+\gamma_2 \frac{\partial \xi_i}{\partial x_3}\Bigg{|}_{x=x(\xi)}+
(\gamma_1 -2\gamma_3 \alpha)\frac{\partial \xi_i}{\partial x_2}\Bigg{|}_{x=x(\xi)}\\
& {\hspace{10cm}{ i=2,3,4}}\\
\{\pi_2,\pi_3\}=&-\frac{\partial \beta_1}{\partial x_2}\Bigg{|}_{x=x(\xi)} \pi_2 +  \left( \frac{\partial \alpha}{\partial x_3}+\frac{\partial }{\partial x_2}(\alpha \beta_1-\beta_0) \right)\Bigg{|}_{x=x(\xi)}\\
\{\pi_2,\pi_4\}=& \frac{\partial \gamma_3}{\partial x_2}\Bigg{|}_{x=x(\xi)} {\pi_2}^2-\frac{\partial \gamma_2}{\partial x_2}\Bigg{|}_{x=x(\xi)} \pi_3
+\left( \frac{\partial }{\partial x_2}(2 \alpha \gamma_3 -\gamma_1) + \beta_1 \frac{\partial \gamma_2 }{\partial x_2}\right)\Bigg{|}_{x=x(\xi)}\pi_2 \\
&+\left( \frac{\partial}{\partial x_2} (\alpha \gamma_1-\gamma_0 -\alpha^2 \gamma_3) + (\alpha \beta_1 -\beta_0)\frac{\partial \gamma_2}{\partial x_2}  + \gamma_2 \frac{\partial \alpha}{\partial x_3} +\frac{\partial \alpha}{\partial x_4}\right)\Bigg{|}_{x=x(\xi)}\\
\{\pi_3,\pi_4\}=& \left( 2 \gamma_3 \frac{ \partial \beta_1}{\partial x_2} - \beta_1 \frac{ \partial \gamma_3}{\partial x_2} - \frac{ \partial \gamma_3}{\partial x_3}\right)\Bigg{|}_{x=x(\xi)}{\pi_2}^2 
+\left(-\frac{ \partial \gamma_2}{\partial x_3}-\beta_1\frac{ \partial \gamma_2}{\partial x_2} \right)\Bigg{|}_{x=x(\xi)}\pi_3 \\
&+\left( 2\alpha \frac{ \partial \gamma_3}{\partial x_3}+2 \alpha \beta_1 \frac{ \partial \gamma_3}{\partial x_2} 
-4 \alpha \gamma_3 \frac{ \partial \beta_1}{\partial x_2} + \beta_1 \frac{ \partial \gamma_2}{\partial x_3} +{\beta_1}^2
\frac{ \partial \gamma_2}{\partial x_2}+ 2\gamma_3 \frac{ \partial \beta_0}{\partial x_2}
-\frac{ \partial \gamma_1}{\partial x_3}\right. \\& \left.
 - \beta_1 \frac{ \partial \gamma_1}{\partial x_2} 
+\frac{ \partial \beta_1}{\partial x_4}+ \gamma_2 \frac{ \partial \beta_1}{\partial x_3}
+\gamma_1 \frac{ \partial \beta_1}{\partial x_2}
  \right)\Bigg{|}_{x=x(\xi)}\pi_2 
+\left(-\alpha^2 \frac{ \partial \gamma_3}{\partial x_3} -\alpha^2 \beta_1 \frac{ \partial \gamma_3}{\partial x_2} 
\right.\\&\left.
+2 \alpha^2 \gamma_3 \frac{ \partial \beta_1}{\partial x_2} -\alpha \beta_1 \frac{ \partial \gamma_2}{\partial x_3} + \beta_0 \frac{ \partial \gamma_2}{\partial x_3}- \alpha {\beta_1}^2 \frac{ \partial \gamma_2}{\partial x_2} +
\beta_0 \beta_1 \frac{ \partial \gamma_2}{\partial x_2} -2\alpha \gamma_3  \frac{ \partial \beta_0}{\partial x_2} +\alpha \frac{ \partial \gamma_1}{\partial x_3} \right.\\&\left.
+\alpha \beta_1 \frac{ \partial \gamma_1}{\partial x_2} -\alpha\frac{ \partial \beta_1}{\partial x_4}
-\alpha \gamma_2 \frac{ \partial \beta_1}{\partial x_3} -\alpha \gamma_1 \frac{ \partial \beta_1}{\partial x_2}
-\frac{ \partial \gamma_0}{\partial x_3}+\frac{ \partial \beta_0}{\partial x_4} +\gamma_2 \frac{ \partial \beta_0}{\partial x_3}+ \gamma_1 \frac{ \partial \beta_0}{\partial x_2}
%\right.\\&\left.
-\beta_1 \frac{ \partial \gamma_0}{\partial x_2} 
 \right)\Bigg{|}_{x=x(\xi)}.
 \end{split}
\end{equation}
}
With the choice $\alpha=\frac{1}{3}u_3+\frac{1}{12}(u_4)^2, \beta_0=-\frac{1}{2} u_2,\beta_1=-\frac{1}{2}u_4$ equation (\ref{OO}) takes the canonical form and the Poisson structure (\ref{001}) becomes an appropriate one for constructing three-dimensional deformations of moduli $g_2$ and $g_3$ of the elliptic curves. The corresponding equations are too complicated to be presented here.

 The observations made in this and previous section naturally lead to introduction of a notion of equivalent Poisson structures in the framework of the theory of coisotropic deformations. This remark is due to Jean-Claude Thomas and Volodya Rubtsov. 

\section{Coisotropic deformations of curves and surfaces in $R^{4}$}

Deformations of curves in $R^{4}$ defined by three equations can be studied
analogously to the three-dimensional case. For the curves given by equations

\begin{eqnarray}
f_{1} &=&p_{1}p_{2}+ap_{1}+bp_{2}+c=0, \\
f_{2} &=&p_{3}+\sum_{k=1}^{n}\alpha _{k}p_{1}^{k}=0,\quad
f_{3}=p_{4}+\sum_{k=1}^{m}\beta _{k}p_{2}^{k}=0
\end{eqnarray}
with arbitrary n and m the coisotropic deformations are governed by the CSs
which coincide with equations of the universal Whitham hierarchy on Riemann
sphere with two punctures  (see \cite{KM1}).

\ An interesting particular case corresponds to

\begin{eqnarray}
f_{1} &=&p_{1}p_{2}-1=0, \\
f_{2} &=&p_{3}+ap_{1}-a=0,\quad f_{3}=p_{4}+bp_{2}-b=0
\end{eqnarray}
for which the curve is the intersection of the cylindrical hyperbola and two
hyperplanes. In order to get nontrivial coisotropic deformations in this
case one has to choose the Poisson bracket in the form  (see \cite{KM2})

\begin{equation}
\left\{ f,g\right\} =\sum_{k=1}^{4}\gamma _{k}\left( \frac{\partial f}{%
\partial p_{k}}\frac{\partial g}{\partial x_{k}}-\frac{\partial f}{\partial
x_{k}}\frac{\partial g}{\partial p_{k}}\right)
\end{equation}
where $\gamma _{1}=p_{1},\gamma _{2}=-p_{2},\gamma _{3}=\gamma _{4}=1.$ Then
the coisotropy conditions

\begin{equation}
\left\{ f_{j},f_{k}\right\}\mid_{ \Gamma }=0,\quad j,k=1,2,3
\end{equation}
gives rise to the following CS
\begin{eqnarray}
a_{x_{1}}+a_{x_{2}} &=&0,\quad b_{x_{1}}+b_{x_{2}}=0, \\
a_{x_{4}}+ab_{x_{1}} &=&0,\quad b_{x_{3}}-ba_{x_{2}}=0.
\end{eqnarray}
This system implies that the variable $\Theta =\log (ab)$ obeys the equation
\begin{equation}
\Theta_{x_{3}x_{4}}+( \exp {\Theta} )_{x_{1}x_{1}}=0
\end{equation}
which is well-known Boyer-Finley or dispersionless two-dimensional Toda
lattice (2DTL) equation. Choosing $f_{2}$ and $f_{3}$ as polynomials of
any order in $p_{1}$ and $p_{2}$, respectively, one gets coisotropic
deformations of curves in $R^{4}$ described by higher d2DTL equations.

Coisotropic deformations can be constructed also for algebraic varieties of
other types in $R^{4}$. For instance, let us consider a pencil of
hyperplanes in $R^{4}$ defined by the equations

\begin{eqnarray}
f_{1} &=&p_{3}+(a-\lambda )p_{1}+bp_{2}=0, \\
f_{2} &=&p_{4}+cp_{1}+(d-\lambda )p_{2}=0
\end{eqnarray}
where $\lambda $ is a parameter. The coisotropy condition $\left\{
f_{1},f_{2}\right\} \mid_{ \Gamma }=0$ for all values of $\lambda $ with the
canonical Poisson bracket in $R^{8}$ gives rise to
\begin{equation}
a=\Phi _{x_{1}},\quad b=\widetilde{\Phi }_{x_{1}},\quad c=\Phi
_{x_{2}},\quad d=\widetilde{\Phi }_{x_{2}}
\end{equation}
and the equations

\begin{eqnarray}
\Phi _{x_{1}x_{4}}-\Phi _{x_{2}x_{3}}+\Phi _{x_{2}}\Phi _{x_{1}x_{1}}-\Phi
_{x_{1}}\Phi _{x_{1}x_{2}}+\widetilde{\Phi }_{x_{2}}\Phi _{x_{1}x_{2}}-%
\widetilde{\Phi }_{x_{1}}\Phi _{x_{2}x_{2}} &=&0,  \notag \\
\widetilde{\Phi }_{x_{1}x_{4}}-\widetilde{\Phi }_{x_{2}x_{3}}+\widetilde{%
\Phi }_{x_{2}}\widetilde{\Phi }_{x_{1}x_{2}}-\widetilde{\Phi }_{x_{1}}%
\widetilde{\Phi }_{x_{2}x_{2}}+\Phi _{x_{2}}\widetilde{\Phi }%
_{x_{1}x_{1}}-\Phi _{x_{1}}\widetilde{\Phi }_{x_{1}x_{2}} &=&0.
\end{eqnarray}
This CS describes the coisotropic deformations of the pencil of the
hyperplanes. The system (93) admits the constraint $\Phi =\Theta _{x_{2}},%
\widetilde{\Phi }=-\Theta _{x_{1}}$ under which it is reduced to the single
equation

\begin{equation}
\Theta _{x_{1}x_{4}}-\Theta _{x_{2}x_{3}}+\Theta _{x_{1}x_{1}}\Theta
_{x_{2}x_{2}}-(\Theta _{x_{1}x_{2}})^{2}=\alpha (x_{1},x_{3},x_{4})
\end{equation}%
where $\alpha (x_{1},x_{3},x_{4})$ is an arbitrary function. At $\alpha =0$
it is the celebrated heavenly equation \cite{Pleba}.

This last example has a natural extension to the spaces of any dimension.
Indeed, let us consider a rational pencil of hyperplanes in $R^{n}$ defined
by the equations

\begin{equation}
f_{1}=\sum_{k=1}^{n}a_{k}(\lambda )p_{k}=0,\quad
f_{2}=\sum_{k=1}^{n}b_{k}(\lambda )p_{k}=0
\end{equation}%
where $a_{k}(\lambda )$ and $b_{k}(\lambda )$ are certain rational functions
of the parameter $\lambda $. The coisotropy condition $\left\{
f_{1},f_{2}\right\}\mid_{ \Gamma }=0$ with the canonical Poisson bracket in
$R^{2n}$ for all values of $\lambda $ is equivalent to the system of
differential equations for the coefficients of the rational functions $%
a_{k}(\lambda )$, $b_{k}(\lambda )$. This system coincides with that
considered in \cite{ZS} in connection with the commutativity condition of
multidimensional vector fields. This coincidence is not accidental. It is a
well-known fact that the expression for the commutator of vector fields is
in one-to one correspondence with the expression of the Poisson bracket of
functions linear in momenta $p_{j}$  ($p_{j}\longleftrightarrow \frac{\partial }{\partial x_{j}}$).

\section{Conclusion}

 Coisotropic deformations studied in this paper form a special class of deformations 
of algebraic varieties. CSs describing such deformations represent themselves the differential constraints on coefficients of the functions $f_{j}$, i.e. on the coordinates of parameter space for algebraic variety  (for this notion see e.g. \cite{H}). Solutions of CSs generate particular subvarieties of dimension n  (surfaces, hypersurfaces etc) in the parameter space. Examples of CSs considered in the paper are the integrable hydrodynamical type systems. They have number of remarkable properties  (infinite sets of integrals, symmetries etc). These properties are inherited by the deformed algebraic varieties and they could be of algebro-geometrical relevance.

 For general solutions of CSs each member  (for fixed $x_{i},i=1,...,n$) of the family of deformed varieties is an algebraic variety in $R^{n}$, but the totality of them  (i.e. submanifold $\Gamma$) is not. Polynomial solutions of CSs are then of particular interest. For them the whole family of deformed algebraic varieties is an algebraic variety too (in $R^{2n}$). Thus, for such solutions of CSs families of coisotropically deformed  algebraic varieties belong to the class of families of algebraic varieties which ``vary algebraically with parameters''. This class is ``one that is fundamental in much of algebraic geometry'' \cite{H} (p.41).
 \vspace{1truecm}\\
 \textbf{Acknowledgments} G.O. thanks J.-C.Thomas and V. Rubtsov for the fruitful discussion on the Poisson structures  and A. Previtali for an introduction to the concept of Groebner basis.
 B.G.K. is grateful to the Isaac Newton Institute for Mathematical Sciences  (Cambridge, UK) for 
kind hospitality during the programme ``Discrete Integrable Systems''.

\end{document}